\documentclass[runningheads]{llncs}
\usepackage{fancyhdr}
\usepackage{longtable}
\usepackage{caption}
\usepackage{float}
\usepackage{multirow}
\usepackage{amsmath}
\usepackage{amssymb}

\usepackage[%
  paper=letterpaper,      
  left=1.5in,            
  right=1.5in,           
  top=1.5in,                
  bottom=1.5in
]{geometry}
\usepackage{graphicx}
\usepackage{lipsum}
\usepackage{url}
\usepackage{booktabs}

\begin{document}
\title{optHIM: Hybrid Iterative Methods for Continuous Optimization in PyTorch}
\author{Nikhil Sridhar and Sajiv Shah}

\maketitle

\begin{figure}[ht]
    \centering
    \begin{minipage}{0.45\textwidth}
        \centering
        \includegraphics[width=\linewidth]{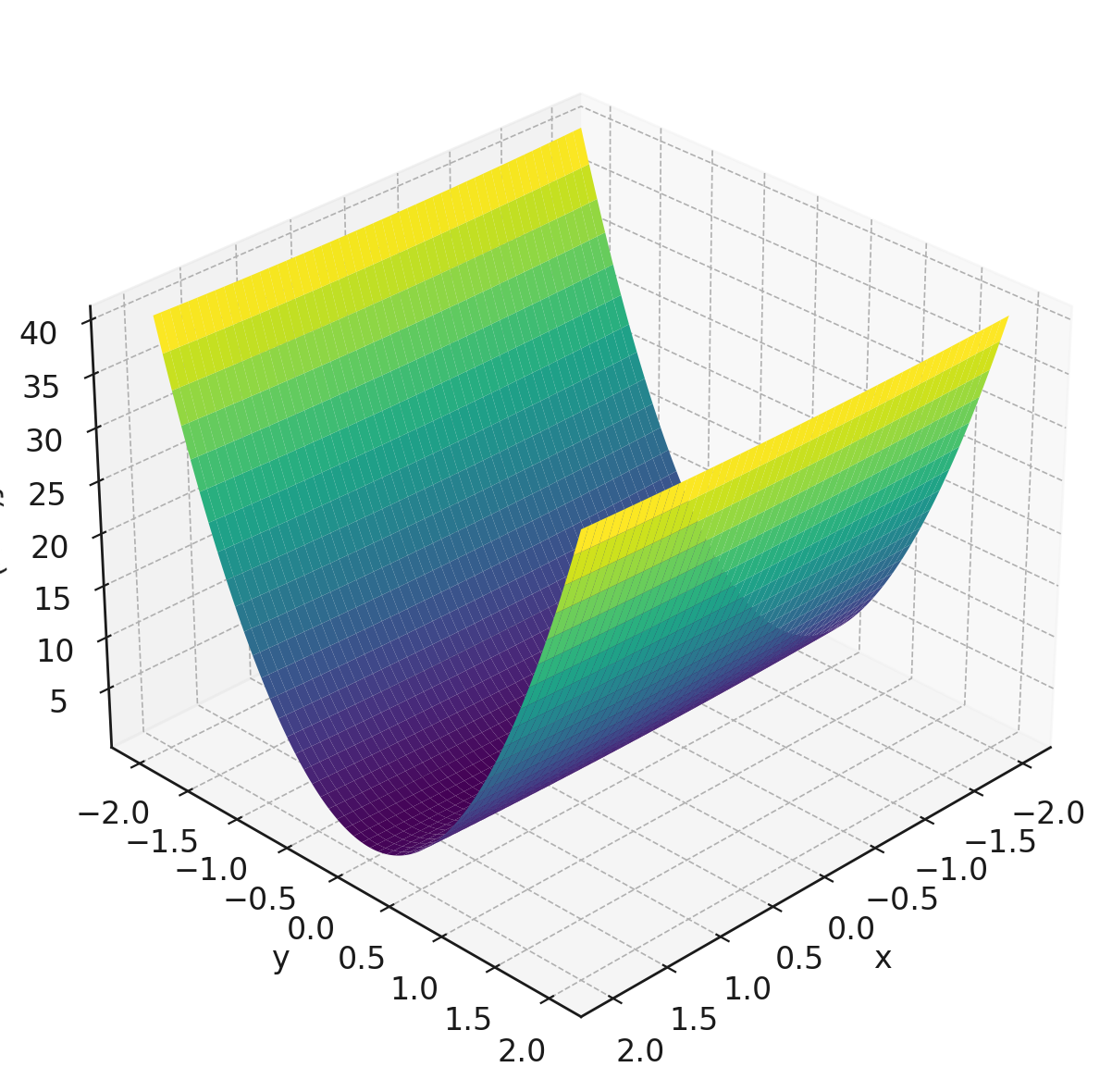}
        \caption*{(a) Simple Quadratic}
    \end{minipage}
    \hfill
    \begin{minipage}{0.45\textwidth}
        \centering
        \includegraphics[width=\linewidth]{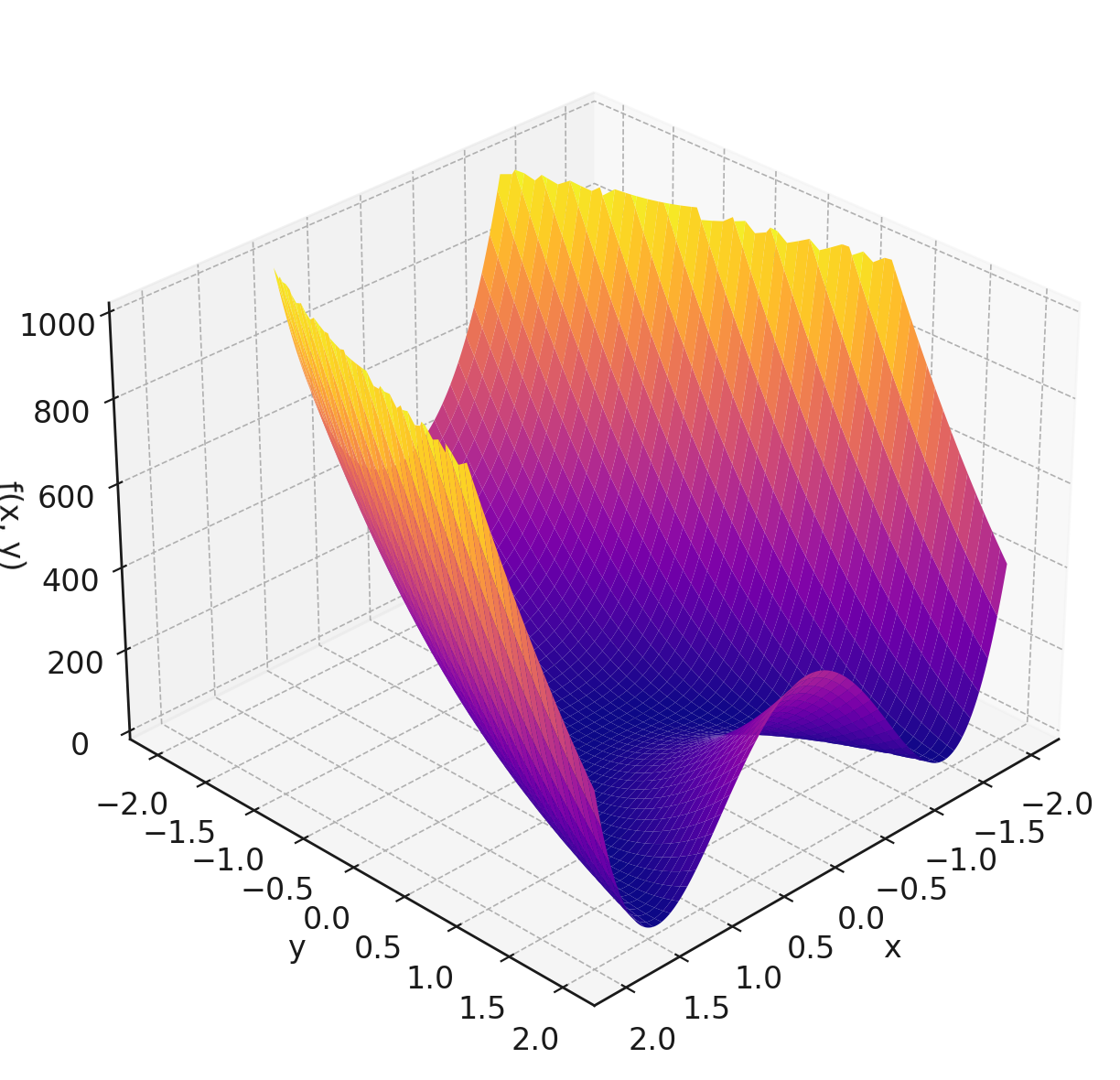}
        \caption*{(b) Rosen\_A}
    \end{minipage}
    \caption{\textbf{Function visualization.} Two representative functions from our benchmark suite are shown. The quadratic function, which is ill-conditioned, presents challenges due to its high condition number. The Rosenbrock function is defined by a long, curved valley whose flat base and steep sides produce extreme variation in curvature, making it notoriously difficult to optimize.}
    \label{fig:3d}
\end{figure}

\vspace{-0.5cm}

\begin{abstract}
We introduce \textbf{optHIM}, an open-source library of continuous unconstrained optimization algorithms implemented in PyTorch for both CPU and GPU. By leveraging PyTorch’s autograd, optHIM seamlessly integrates function, gradient, and Hessian information into flexible line search and trust region methods. We evaluate eleven state-of-the-art variants on benchmark problems spanning convex and non-convex landscapes. Through a suite of quantitative metrics and qualitative analyses, we demonstrate each method’s strengths and trade-offs. optHIM aims to democratize advanced optimization by providing a transparent, extensible, and efficient framework for research and education.
\end{abstract}

\section{Algorithm Overview}

\subsection{Line Search Methods}
We implement the five line search methods below for minimizing \(f\colon\mathbb{R}^n\to\mathbb{R}\).  Each update has the form
\[
x_{k+1} = x_k + \alpha_k\,p_k
\tag{1}
\]
where \(\alpha_k\) is the step size and \(p_k\) is the direction at iteration \(k\).

\subsubsection{Gradient Descent (GD)} sets the search direction as \(p_k = -\nabla f(x_k)\).

\subsubsection{Newton’s Method} uses the exact Hessian to compute
\[
p_k = -\nabla^2f(x_k)^{-1}\nabla f(x_k)
\tag{1}
\]
If $\nabla^2f(x_k)$ is not positive definite, inversion may not be possible. In our implementation, we iteratively correct the Hessian until it is invertible by adding factors of the identity matrix. All together, this method incurs an \(\mathcal{O}(n^3)\) computational cost for each iteration and \(\mathcal{O}(n^2)\) storage cost to save the Hessian matrix. 

\bigskip

\noindent The quasi‐Newton methods below approximate the Hessian as $B_k$ by enforcing the secant equation
\[
B_{k+1}\,s_k = y_k,\quad s_k = x_{k+1}-x_k,\quad y_k = \nabla f(x_{k+1})-\nabla f(x_k)
\tag{2}
\]
This ensures that the gradient of the model matches the true gradient at both \(x_k\) and \(x_{k+1}\)

\subsubsection{Broyden–Fletcher–Goldfarb–Shanno (BFGS) \cite{FletcherPowell1963} }approximates $B_k^{-1} = H_k$ with the update
\[
H_{k+1} =
\Bigl(I - \tfrac{s_k y_k^T}{y_k^T s_k}\Bigr)\,H_k\,
\Bigl(I - \tfrac{y_k s_k^T}{y_k^T s_k}\Bigr)
+ \tfrac{s_k s_k^T}{y_k^T s_k}
\tag{3}
\]

\subsubsection{Davidon–Fletcher–Powell (DFP) \cite{Davidon1959}} updates $H_k$ according to
\[
H_{k+1} = H_k + \frac{s_k s_k^T}{s_k^T y_k}
    - \frac{H_k\,y_k\,y_k^T\,H_k}{y_k^T H_k y_k}
\tag{4}
\]

\noindent BFGS and DFP both are symmetric rank 2 updates. They incur an \(\mathcal{O}(n^2)\) computational cost for each iteration and \(\mathcal{O}(n^2)\) storage cost to save $H_k$.  They preserve positive‐definiteness provided the condition
\[
y_k^T s_k > 0
\tag{5}
\]
Thus, we skip the update if $|y_k^Ts_k| \le \epsilon_{sy}\,||y_k||\,||s_k||$ for $\epsilon_{sy} = 1e^{-6}$. 

\subsubsection{Limited‐Memory BFGS (L-BFGS) \cite{Nocedal1980}} retains only the most recent \(m\) pairs \((s_k,y_k)\). This reduces both time and memory complexity to \(\mathcal{O}(m n)\), making it well-suited for large‐scale problems while preserving convergence behavior similar to BFGS.

\paragraph{Backtracking Line Search}  
We begin with an initial step size \(\alpha_{\mathrm{init}}\) and iteratively reduce \(\alpha\leftarrow\tau\,\alpha\) until the Armijo condition below is satisfied.  
\[
f(x_k + \alpha\,p_k)\;\le\;f(x_k) + c_1\,\alpha\,\nabla f(x_k)^T p_k
\tag{6}
\]  

\noindent For Wolfe backtracking, we then additionally require the curvature condition  
\[
\nabla f(x_k + \alpha\,p_k)^T p_k \;\ge\; c_2\,\nabla f(x_k)^T p_k
\tag{7}
\]  

\noindent Parameters for line search methods are defined in Table~\ref{tab:ls-params}.  

\begin{figure}[ht]
    \centering
    \begin{minipage}{0.48\textwidth}
        \centering
        \includegraphics[width=\linewidth]{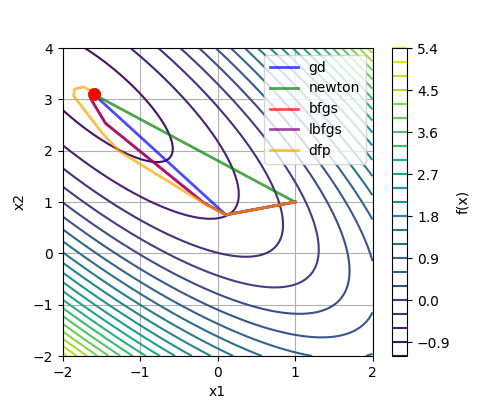}
        \caption*{(a) Simple Quadratic}
    \end{minipage}
    \hfill
    \begin{minipage}{0.48\textwidth}
        \centering
        \includegraphics[width=\linewidth]{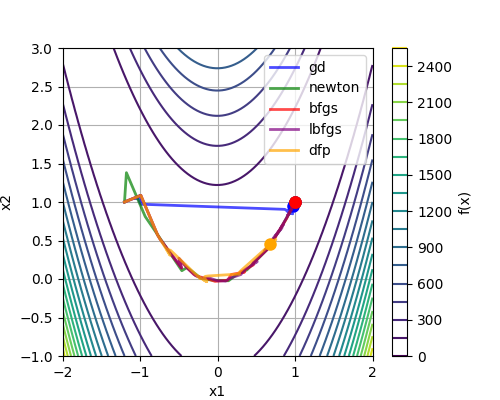}
        \caption*{(b) Rosen\_A}
    \end{minipage}
    \caption{\textbf{Line search trajectory comparison.} Trajectories of line search algorithms from Table~\ref{tab:ls_metrics} on 3D quadratic and Rosenbrock problems. The solution is marked by a bright red circle, and each algorithm’s final point is shown as a colored circle matching its trajectory. For the simple quadratic, the initial point is \((1, 1)\). For the Rosenbrock problem, the initial point is randomized within a small neighborhood of \((-1, 1)\). For each algorithm, only the variant (Armijo or Wolfe) that achieved the better performance on the problem was selected for inclusion in the plot.}
    \label{fig:ls_traj}
\end{figure}

\begin{figure}[ht]
    \centering
    \begin{minipage}{0.48\textwidth}
        \centering
        \includegraphics[width=\linewidth]{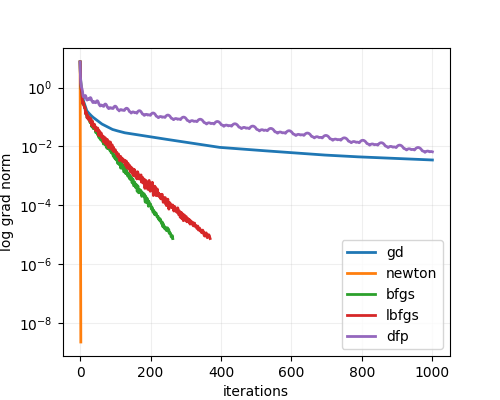}
        \caption*{(a) Quad\_D}
    \end{minipage}
    \hfill
    \begin{minipage}{0.48\textwidth}
        \centering
        \includegraphics[width=\linewidth]{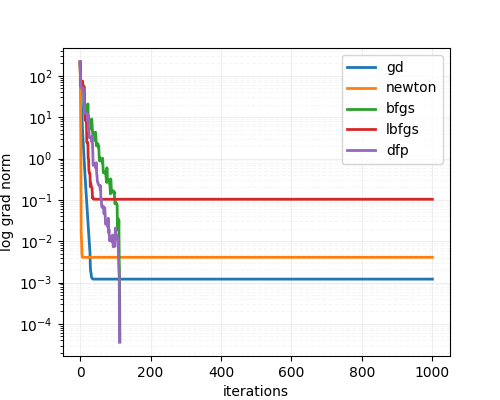}
        \caption*{(b) Rosen\_B}
    \end{minipage}
    \caption{\textbf{Line search convergence comparison.} Convergence profiles of line search algorithms from Table~\ref{tab:ls_metrics} on high-dimensional quadratic and Rosenbrock problems. The plots show the logarithm of the gradient norm (a measure of stationarity) versus the number of iterations. For each algorithm, only the variant (Armijo or Wolfe) that achieved the better performance on the problem was selected for inclusion in the plot.}
    \label{fig:ls_gn}
\end{figure}

\subsection{Trust Region Methods}

\subsubsection{Models}

Trust region methods build and minimize the quadratic model
\begin{equation}
m_k(p) = f(x_k) + \nabla f(x_k)^T p + \tfrac12\,p^T B_k p,
\tag{8}
\end{equation}
subject to \(\|p\|\le \delta_k\).  We consider four variants for \(B_k\), three borrowed from line search methods (Newton, BFGS, DFP) and one simpler update below.

\paragraph{Symmetric Rank‐One (SR1)}  
The SR1 update \cite{ConnGouldToint2000} defines
\begin{equation}
B_{k+1} = B_k + \frac{(y_k - B_k\,s_k)(y_k - B_k\,s_k)^T}{(y_k - B_k\,s_k)^T s_k},
\tag{9}
\end{equation}
We skip this update when \(|(y_k - B_k s_k)^T s_k| < c_3\,||(y_k - B_k s_k)||\,||s_k||\) to maintain numerical stability. The SR1 update does not guarantee positive definiteness. 

\subsubsection{Subproblem Solvers}

\paragraph{Cauchy Step}  
The Cauchy step uses the model gradient to define
\begin{equation}
p_k = -\alpha_C\,\nabla f(x_k),
\quad
\alpha_C = \min\!\Bigl\{\frac{\|\nabla f(x_k)\|^2}{\nabla f(x_k)^T B_k \,\nabla f(x_k)},\,\frac{\delta_k}{\|\nabla f(x_k)\|}\Bigr\}.
\tag{10}
\end{equation}

\paragraph{Truncated Conjugate Gradient (CG)}  
The CG solver approximately solves \(B_k p = -\nabla f(x_k)\).  Iterations stop when \(\|p\|\) reaches \(\delta_k\) or when negative curvature is detected.  We limit CG to \emph{max\_iter} steps and require the residual norm to fall below \emph{tol}.

\paragraph{Radius Update}  
After computing \(p_k\), we evaluate the ratio
\begin{equation}
\rho_k = \frac{f(x_k) - f(x_k + p_k)}{m_k(0) - m_k(p_k)}.
\tag{11}
\end{equation}
We then update the radius by
\begin{equation}
\delta_{k+1} =
\begin{cases}
\frac{1}{2}\,\delta_k, & \rho_k < c_1,\\
2\,\delta_k, & \rho_k > c_2,\\
\delta_k, & \text{otherwise},
\end{cases}
\tag{12}
\end{equation}
Parameters for trust region methods are defined in Table~\ref{tab:tr-params}.

\section{Implementation}
Our algorithms are implemented in PyTorch by defining only each objective’s \texttt{forward} method.  PyTorch’s autograd automatically computes gradients and Hessian–vector products, so we avoid manual derivative code.  

We wrap PyTorch’s \texttt{Optimizer} API to inject custom line search and trust region logic.  In line search, the step size is adapted via Armijo/Wolfe backtracking on the loss returned by \texttt{forward}.  In trust region, we reuse the same backtracking routines to build quadratic models and solve subproblems with Cauchy or CG steps.

The optHIM repository exposes a single configuration object for each run.  Users can specify the algorithm (e.g.\ DFP model with CG solver), the benchmark function, stopping criteria, maximum iterations, and all line search or trust region parameters.  This design makes it trivial to reproduce experiments or explore new variants by editing a YAML file rather than source code.

\begin{longtable}{@{}llccccc@{}}
\label{tab:ls_metrics} \\
\toprule
Problem & Metric & GD & Newton & BFGS & L-BFGS & DFP \\
\midrule
\endfirsthead

\multicolumn{7}{c}{\tablename\ \thetable{} — continued}\\
\toprule
Problem & Metric & GD & Newton & BFGS & L-BFGS & DFP \\
\midrule
\endhead

\midrule \multicolumn{7}{r}{Continued on next page} \\
\endfoot

\bottomrule \\
\caption{\textbf{Line search evaluation.} Performance of line search algorithms across 11 problems of varying geometry and dimension. Each entry reports results for the method using backtracking line search with the Armijo \textbar{} Wolfe conditions. The best value for each metric across both variants is bolded. Runtimes were measured on CPU.}
\endlastfoot

\multirow{5}{*}{Quad\_A}
  & Iterations  & 98 \textbar{} 98 & \textbf{1} \textbar{} \textbf{1} & 25 \textbar{} 25 & 29 \textbar{} 29 & 38 \textbar{} 38 \\
  & Func Evals  & 295 \textbar{} 295 & \textbf{4} \textbar{} \textbf{4} & 76 \textbar{} 76 & 88 \textbar{} 88 & 115 \textbar{} 115 \\
  & Grad Evals  & 99 \textbar{} 197 & \textbf{2} \textbar{} 3 & 26 \textbar{} 51 & 30 \textbar{} 59 & 39 \textbar{} 77 \\
  & Time (s)    & 0.01 \textbar{} 0.02 & \textbf{0.00} \textbar{} 0.01 & 0.01 \textbar{} 0.01 & 0.01 \textbar{} 0.01 & 0.01 \textbar{} 0.01 \\
  & Converged?  & T \textbar{} T & T \textbar{} T & T \textbar{} T & T \textbar{} T & T \textbar{} T \\

\midrule
\multirow{5}{*}{Quad\_B}
  & Iterations  & 1000 \textbar{} 1000 & \textbf{2} \textbar{} \textbf{2} & 57 \textbar{} 57 & 181 \textbar{} 181 & 1000 \textbar{} 1000 \\
  & Func Evals  & 3001 \textbar{} 4171 & \textbf{7} \textbar{} \textbf{7} & 172 \textbar{} 172 & 544 \textbar{} 544 & 3001 \textbar{} 3001 \\
  & Grad Evals  & 1001 \textbar{} 2196 & \textbf{3} \textbar{} 5 & 58 \textbar{} 115 & 182 \textbar{} 363 & 1001 \textbar{} 2001 \\
  & Time (s)    & 0.13 \textbar{} 0.25 & \textbf{0.00} \textbar{} \textbf{0.00} & 0.01 \textbar{} 0.01 & 0.03 \textbar{} 0.04 & 0.15 \textbar{} 0.20 \\
  & Converged?  & F \textbar{} F & T \textbar{} T & T \textbar{} T & T \textbar{} T & F \textbar{} F \\

\midrule
\multirow{5}{*}{Quad\_C}
  & Iterations  & 108 \textbar{} 108 & \textbf{2} \textbar{} \textbf{2} & 30 \textbar{} 30 & 31 \textbar{} 31 & 42 \textbar{} 42 \\
  & Func Evals  & 325 \textbar{} 325 & \textbf{7} \textbar{} \textbf{7} & 91 \textbar{} 91 & 94 \textbar{} 94 & 127 \textbar{} 127 \\
  & Grad Evals  & 109 \textbar{} 217 & \textbf{3} \textbar{} 5 & 31 \textbar{} 61 & 32 \textbar{} 63 & 43 \textbar{} 85 \\
  & Time (s)    & 0.02 \textbar{} 0.04 & 0.22 \textbar{} 0.25 & 0.14 \textbar{} 0.14 & \textbf{0.01} \textbar{} 0.02 & 0.18 \textbar{} 0.19 \\
  & Converged?  & T \textbar{} T & T \textbar{} T & T \textbar{} T & T \textbar{} T & T \textbar{} T \\

\midrule
\multirow{5}{*}{Quad\_D}
  & Iterations  & 1000 \textbar{} 1000 & \textbf{2} \textbar{} \textbf{2} & 263 \textbar{} 263 & 369 \textbar{} 369 & 1000 \textbar{} 1000 \\
  & Func Evals  & 3001 \textbar{} 4165 & \textbf{7} \textbar{} \textbf{7} & 790 \textbar{} 790 & 1108 \textbar{} 1108 & 3001 \textbar{} 3001 \\
  & Grad Evals  & 1001 \textbar{} 2195 & \textbf{3} \textbar{} 5 & 264 \textbar{} 527 & 370 \textbar{} 739 & 1001 \textbar{} 2001 \\
  & Time (s)    & 0.22 \textbar{} 0.41 & 0.22 \textbar{} 0.22 & 1.28 \textbar{} 1.32 & 0.19 \textbar{} \textbf{0.18} & 4.79 \textbar{} 5.26 \\
  & Converged?  & F \textbar{} F & T \textbar{} T & T \textbar{} T & T \textbar{} T & F \textbar{} F \\

\midrule
\multirow{5}{*}{Quartic\_A}
  & Iterations  & \textbf{2} \textbar{} \textbf{2} & \textbf{2} \textbar{} \textbf{2} & 3 \textbar{} 3 & 3 \textbar{} 3 & 3 \textbar{} 3 \\
  & Func Evals  & \textbf{7} \textbar{} \textbf{7} & \textbf{7} \textbar{} \textbf{7} & 10 \textbar{} 10 & 10 \textbar{} 10 & 10 \textbar{} 10 \\
  & Grad Evals  & \textbf{3} \textbar{} 5 & \textbf{3} \textbar{} 5 & 4 \textbar{} 7 & 4 \textbar{} 7 & 4 \textbar{} 7 \\
  & Time (s)    & \textbf{0.00} \textbar{} \textbf{0.00} & \textbf{0.00} \textbar{} \textbf{0.00} & \textbf{0.00} \textbar{} \textbf{0.00} & \textbf{0.00} \textbar{} \textbf{0.00} & \textbf{0.00} \textbar{} \textbf{0.00} \\
  & Converged?  & T \textbar{} T & T \textbar{} T & T \textbar{} T & T \textbar{} T & T \textbar{} T \\

\midrule
\multirow{5}{*}{Quartic\_B}
  & Iterations  & \textbf{6} \textbar{} \textbf{6} & 12 \textbar{} 12 & 25 \textbar{} 25 & 18 \textbar{} 18 & 68 \textbar{} 68 \\
  & Func Evals  & 100 \textbar{} 100 & \textbf{37} \textbar{} \textbf{37} & 138 \textbar{} 138 & 171 \textbar{} 171 & 266 \textbar{} 266 \\
  & Grad Evals  & \textbf{7} \textbar{} 13 & 13 \textbar{} 25 & 26 \textbar{} 51 & 19 \textbar{} 37 & 69 \textbar{} 137 \\
  & Time (s)    & \textbf{0.00} \textbar{} \textbf{0.00} & 0.01 \textbar{} 0.01 & \textbf{0.00} \textbar{} 0.01 & \textbf{0.00} \textbar{} 0.01 & 0.01 \textbar{} 0.01 \\
  & Converged?  & T \textbar{} T & T \textbar{} T & T \textbar{} T & T \textbar{} T & T \textbar{} T \\

\midrule
\multirow{5}{*}{Rosen\_A}
  & Iterations  & 1000 \textbar{} 1000 & \textbf{20} \textbar{} \textbf{20} & 34 \textbar{} 34 & 34 \textbar{} 34 & 1000 \textbar{} 860 \\
  & Func Evals  & 11835 \textbar{} 11883 & \textbf{68} \textbar{} \textbf{68} & 121 \textbar{} 121 & 133 \textbar{} 133 & 9641 \textbar{} 2669 \\
  & Grad Evals  & 1001 \textbar{} 2002 & \textbf{21} \textbar{} 41 & 35 \textbar{} 69 & 35 \textbar{} 69 & 1001 \textbar{} 1735 \\
  & Time (s)    & 0.25 \textbar{} 0.31 & \textbf{0.01} \textbar{} \textbf{0.01} & \textbf{0.01} \textbar{} \textbf{0.01} & \textbf{0.01} \textbar{} \textbf{0.01} & 0.23 \textbar{} 0.16 \\
  & Converged?  & F \textbar{} F & T \textbar{} T & T \textbar{} T & T \textbar{} T & F \textbar{} T \\

\midrule
\newpage
\multirow{5}{*}{Rosen\_B}
  & Iterations  & 1000 \textbar{} 1000 & 1000 \textbar{} 1000 & 113 \textbar{} 113 & 1000 \textbar{} 1000 & 1000 \textbar{} \textbf{112} \\
  & Func Evals  & 18805 \textbar{} 99062 & 10955 \textbar{} 101412 & 1242 \textbar{} 1242 & 19708 \textbar{} 94694 & 7083 \textbar{} \textbf{1038} \\
  & Grad Evals  & 1001 \textbar{} 46481 & 1001 \textbar{} 50708 & \textbf{114} \textbar{} 227 & 1001 \textbar{} 38723 & 1001 \textbar{} 304 \\
  & Time (s)    & 0.37 \textbar{} 4.43 & 6.71 \textbar{} 11.73 & \textbf{0.03} \textbar{} 0.04 & 0.45 \textbar{} 4.00 & 0.19 \textbar{} 0.04 \\
  & Converged?  & F \textbar{} F & F \textbar{} F & T \textbar{} T & F \textbar{} F & F \textbar{} T \\

\midrule
\multirow{5}{*}{Exp\_A}
  & Iterations  & 1000 \textbar{} 29 & \textbf{13} \textbar{} \textbf{13} & 14 \textbar{} 17 & 1000 \textbar{} 23 & 1000 \textbar{} 1000 \\
  & Func Evals  & 12762 \textbar{} 300 & \textbf{57} \textbar{} \textbf{57} & 65 \textbar{} 360 & 9897 \textbar{} 483 & 12876 \textbar{} 101417 \\
  & Grad Evals  & 1001 \textbar{} 150 & \textbf{14} \textbar{} 27 & 15 \textbar{} 187 & 1001 \textbar{} 245 & 1001 \textbar{} 51702 \\
  & Time (s)    & 0.34 \textbar{} 0.02 & 0.01 \textbar{} 0.01 & \textbf{0.00} \textbar{} 0.02 & 0.31 \textbar{} 0.03 & 0.35 \textbar{} 5.63 \\
  & Converged?  & F \textbar{} T & T \textbar{} T & T \textbar{} T & F \textbar{} T & F \textbar{} F \\

\midrule
\multirow{5}{*}{Exp\_B}
  & Iterations  & 1000 \textbar{} 29 & \textbf{13} \textbar{} \textbf{13} & 14 \textbar{} 24 & 17 \textbar{} 19 & 1000 \textbar{} 75 \\
  & Func Evals  & 12762 \textbar{} 300 & \textbf{57} \textbar{} \textbf{57} & 65 \textbar{} 485 & 58 \textbar{} 278 & 7875 \textbar{} 3052 \\
  & Grad Evals  & 1001 \textbar{} 151 & \textbf{14} \textbar{} 27 & 15 \textbar{} 233 & 18 \textbar{} 131 & 1001 \textbar{} 1436 \\
  & Time (s)    & 0.31 \textbar{} 0.02 & 0.01 \textbar{} 0.01 & \textbf{0.00} \textbar{} 0.03 & \textbf{0.00} \textbar{} 0.02 & 0.24 \textbar{} 0.16 \\
  & Converged?  & F \textbar{} T & T \textbar{} T & T \textbar{} T & T \textbar{} T & F \textbar{} T \\

\midrule
\multirow{5}{*}{Genhumps}
  & Iterations  & 175 \textbar{} 124 & 1000 \textbar{} 1000 & 46 \textbar{} 47 & 37 \textbar{} \textbf{26} & 1000 \textbar{} 496 \\
  & Func Evals  & 731 \textbar{} 519 & 30926 \textbar{} 25934 & 155 \textbar{} 164 & 130 \textbar{} \textbf{118} & 3023 \textbar{} 1535 \\
  & Grad Evals  & 176 \textbar{} 252 & 1001 \textbar{} 2002 & 47 \textbar{} 96 & \textbf{38} \textbar{} 57 & 1001 \textbar{} 998 \\
  & Time (s)    & 0.08 \textbar{} 0.08 & 3.91 \textbar{} 3.81 & \textbf{0.02} \textbar{} 0.03 & \textbf{0.02} \textbar{} \textbf{0.02} & 0.36 \textbar{} 0.29 \\
  & Converged?  & T \textbar{} T & F \textbar{} F & T \textbar{} T & T \textbar{} T & F \textbar{} T \\

\end{longtable}

\section{Experiments}

\subsection{Benchmark Functions}
Our evaluation suite comprises eleven functions with diverse geometry:

\paragraph{Quadratic\_A–D}  
Non-convex quadratics of increasing dimension (10 to 1000) and worsening condition number.

\paragraph{Quartic\_A, Quartic\_B}  
Fourth-order polynomials featuring multiple local minima.

\paragraph{Rosen\_A, Rosen\_B}  
The classic 3-dimensional Rosenbrock and its 100-dimensional extension.

\paragraph{Exp\_A, Exp\_B}  
A smooth exponential-quartic hybrid:
\begin{equation}
f_{\mathrm{Exp\_A}}(x) = \frac{e^{x_0}-1}{e^{x_0}+1} + 0.1\,e^{-x_0}
  + \sum_{i=1}^{9}(x_i - 1)^4,
\tag{13}
\end{equation}
with Exp\_B its 100-dimensional analogue.

\paragraph{Genhumps}  
A 5-dimensional “generalized humps” function:
\begin{equation}
f_{\mathrm{Genhumps}}(x) = 
  \sum_{i=1}^{4}\Bigl[\sin^2(2x_{i-1})\,\sin^2(2x_i)
  + 0.05\,(x_{i-1}^2 + x_i^2)\Bigr].
\tag{14}
\end{equation}

The quadratic functions range from mildly to severely ill-conditioned. The quartic and Genhumps functions exhibit pronounced non-convexity. The Rosenbrock problems feature narrow, curved valleys. The exponential hybrids combine steep and flat regions.  

\subsection{Evaluation Protocol}
We terminate each run when 
\begin{equation}
\|\nabla f(x)\|\le10^{-6}
\quad\text{or}\quad
k\ge1000.
\tag{15}
\end{equation}
At termination we record the number of iterations, function and gradient evaluations, CPU time, and a convergence flag.  Summary metrics appear in Tables~\ref{tab:ls_metrics} and \ref{tab:tr_metrics}.

\begin{figure}[ht]
    \centering
    \begin{minipage}{0.48\textwidth}
        \centering
        \includegraphics[width=\linewidth]{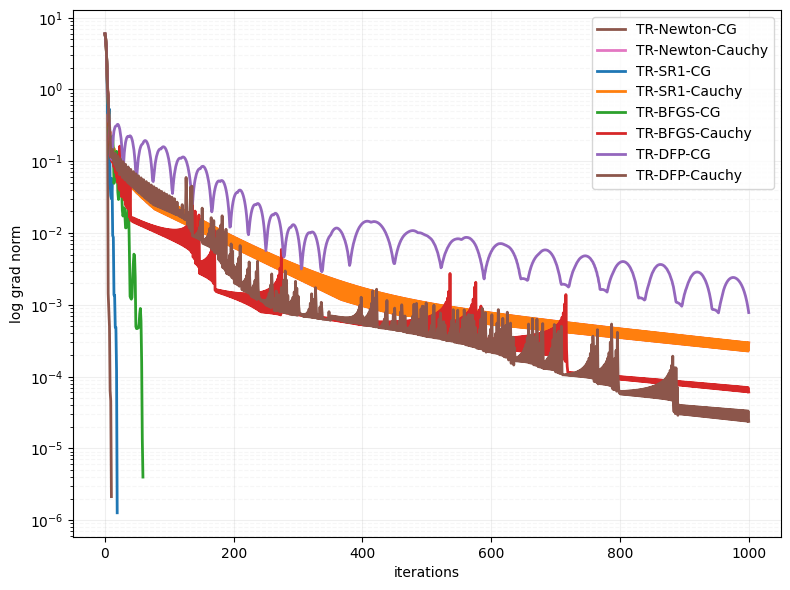}
        \caption*{(a) Quad\_B}
    \end{minipage}
    \hfill
    \begin{minipage}{0.48\textwidth}
        \centering
        \includegraphics[width=\linewidth]{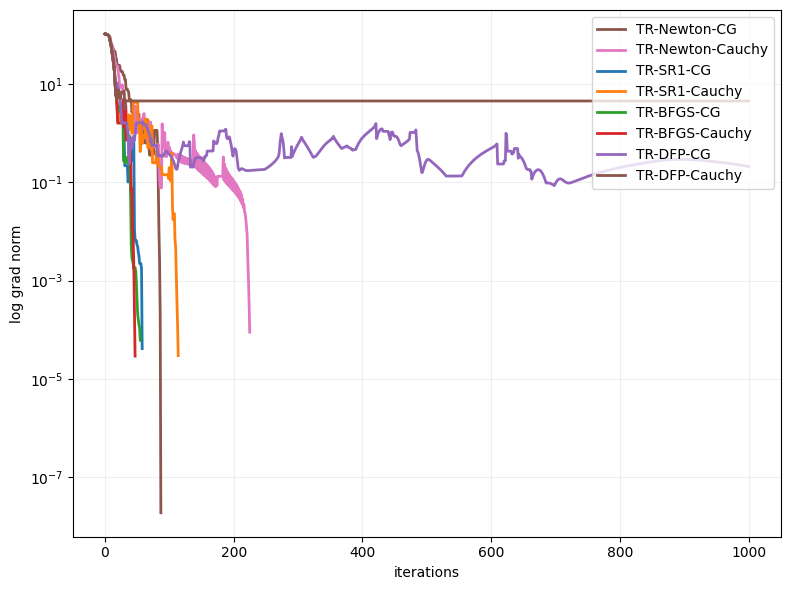}
        \caption*{(b) Genhumps}
    \end{minipage}
    \caption{\textbf{Trust region convergence comparison.} Convergence profiles of trust region algorithms from Table~\ref{tab:tr_metrics} on quadratic and Genhumps problems. The plots show the logarithm of the gradient norm (a measure of stationarity) versus the number of iterations.}
    \label{fig:tr_gn}
\end{figure}

\paragraph{Stationarity Profiles}  
We plot \(\log\|\nabla f(x)\|\) versus iteration number to assess stationarity without a known optimum.  Line search profiles are shown in Figure~\ref{fig:ls_gn}, and trust region profiles in Figure~\ref{fig:tr_gn}.

\begin{figure}[ht]
    \centering
    \begin{minipage}{0.48\textwidth}
        \centering
        \includegraphics[width=\linewidth]{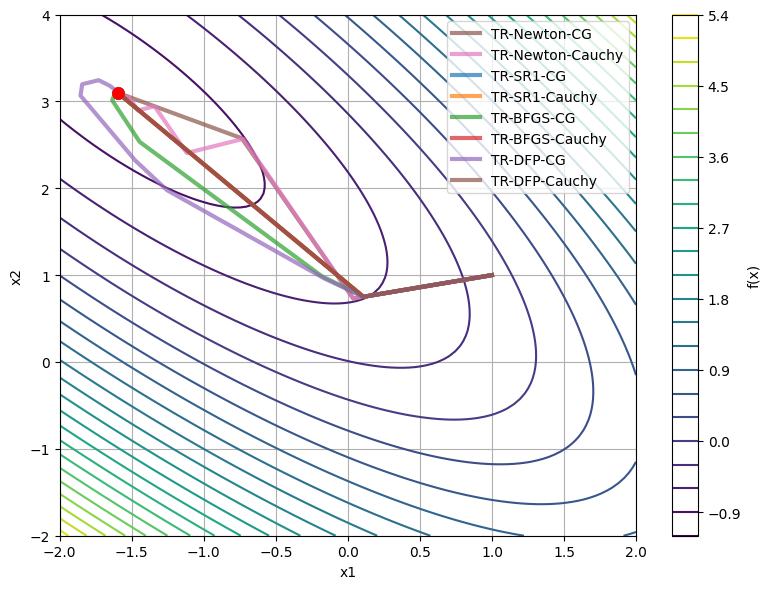}
        \caption*{(a) Simple Quadratic}
    \end{minipage}
    \hfill
    \begin{minipage}{0.48\textwidth}
        \centering
        \includegraphics[width=\linewidth]{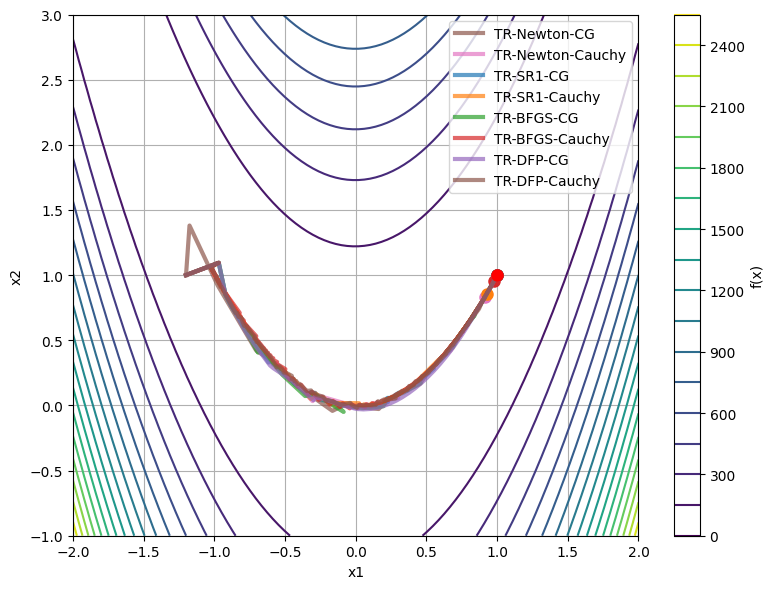}
        \caption*{(b) Rosen\_A}
    \end{minipage}
    \caption{\textbf{Trust region trajectory comparison.} Trajectories of trust region algorithms from Table~\ref{tab:tr_metrics} on 3D quadratic and Rosenbrock problems. The solution is marked by a bright red circle, and each algorithm’s final point is shown as a colored circle matching its trajectory. For the simple quadratic, the initial point is \((1, 1)\). For the Rosenbrock problem, the initial point is randomized within a small neighborhood of \((-1, 1)\).}
    \label{fig:tr_traj}
\end{figure}

\paragraph{Trajectory Comparisons}  
For 3D problems, we overlay algorithmic paths on contour maps.  Figures~\ref{fig:ls_traj} and \ref{fig:tr_traj} illustrate how each method navigates narrow valleys and ill-conditioned basins.

\paragraph{Summary Tables}  
Table~\ref{tab:ls_metrics} reports line search performance across all benchmarks, bolding the best metric per problem.  Table~\ref{tab:tr_metrics} provides the analogous results for trust region variants.  

\begin{table}[ht]
  \centering
  \captionsetup{skip=1em}  
  \begin{tabular}{@{}lccccccc@{}}
    \toprule
    Parameter           & $\alpha_{\mathrm{init}}$ & $\alpha_{\mathrm{low}}$ & $\alpha_{\mathrm{high}}$ & $\tau$ & $c_1$     & $c_2$ & $c$   \\
    \midrule
    Value               & 1.0                      & 0.0                      & 1000.0                    & 0.5     & $10^{-4}$ & 0.9   & 0.5   \\
    \bottomrule
  \end{tabular}
  \caption{\textbf{Line search parameters:} initial, lower, and upper bounds for step size ($\alpha$); backtracking factor ($\tau$); Armijo/Wolfe constants ($c_1$, $c_2$); and interpolation parameter ($c$).}
  \label{tab:ls-params}
\end{table}

\begin{table}[ht]
  \centering
  \captionsetup{skip=1em}
  \begin{tabular}{@{}lcccccccc@{}}
    \toprule
    Parameter     & $\delta_0$ & $\delta_{\min}$ & $\delta_{\max}$ & $c_1$   & $c_2$   & $c_3$     & tol       & max\_iter \\
    \midrule
    Value         & 1.0        & $10^{-6}$       & $10^{2}$        & 0.25    & 0.75    & $10^{-6}$ & $10^{-6}$ & 10        \\
    \bottomrule
  \end{tabular}
  \caption{\textbf{Trust region parameters:} initial, minimum, and maximum radii ($\delta$); acceptance thresholds ($c_1$, $c_2$); SR1‐skip threshold ($c_3$); CG tolerance; and maximum CG iterations.}
  \label{tab:tr-params}
\end{table}

\section{Analysis}

The data in Tables~\ref{tab:ls_metrics} and \ref{tab:tr_metrics} reveal that a low iteration count does not always imply the fastest runtime.  For example, Newton’s method converges in one or two steps on many problems (see Quad\_A–D) but still incurs substantial CPU time when forming and factorizing the Hessian.  In contrast, L-BFGS typically requires more iterations than Newton and BFGS but remains competitive in runtime thanks to its \(\mathcal{O}(mn)\) per‐iteration cost.  On highly ill-conditioned quadratics (Quad\_D), L-BFGS outperforms full BFGS in wall‐clock time despite taking more steps.

Among line search methods, BFGS and L-BFGS strike the best balance between iteration count and per‐step cost, whereas DFP often exhibits slower convergence and, in some cases (Quad\_B, Exp\_A), fails to converge within 1000 iterations.  Convergence profiles in Figure~\ref{fig:ls_gn} show that Newton’s method achieves quadratic convergence near the solution—reflected by the steep drop in \(\|\nabla f\|\) after a few iterations—while quasi‐Newton schemes display superlinear convergence once the Hessian approximation becomes accurate.  GD, by contrast, shows only linear convergence, especially visible on ill-conditioned problems.  Trajectory plots in Figure~\ref{fig:ls_traj} further illustrate that DFP’s less accurate curvature can lead to meandering paths, whereas BFGS and L-BFGS pursue more direct routes.

Trust region results tell a similar story.  TR–Newton–CG and TR–Newton–Cauchy require very few iterations (e.g.\ Quad\_A, Table~\ref{tab:tr_metrics}) but pay a high cost per iteration.  The SR1–CG variant often matches Newton in iteration count while reducing runtime, thanks to a cheaper rank‐one update (see Rosen\_A and Rosen\_B).  However, SR1’s lack of a guaranteed positive‐definite model sometimes causes erratic, oscillatory convergence behavior, as seen in Figure~\ref{fig:tr_gn}(b).  BFGS‐based trust region (TR–BFGS–CG) offers a middle ground, combining superlinear convergence with stable runtime.

The convergence curves in Figure~\ref{fig:tr_gn} highlight that CG subproblem solvers typically yield faster reduction in gradient norm than Cauchy steps.  On Genhumps, for example, TR–SR1–CG converges in under 50 iterations with rapid initial progress, whereas Cauchy steps stall and exhibit only linear decay.  Trajectories in Figure~\ref{fig:tr_traj} confirm that CG steps navigate narrow valleys more directly, whereas Cauchy steps sometimes hug the trust‐region boundary before contracting.

Overall, quasi‐Newton approaches achieve superlinear convergence once sufficient curvature information is captured, while Newton methods demonstrate local quadratic rates at the expense of higher per‐step cost.  Gradient descent maintains only linear convergence, making it less suitable for stiff or ill-conditioned problems.  These trends emphasize the trade‐off between per‐iteration complexity and asymptotic convergence rate across different problem geometries.  

\setlength{\tabcolsep}{6pt}

\begin{longtable}{@{}ll|c|c|c|c|c|c|c|c@{}}
\label{tab:tr_metrics} \\
\toprule
\multirow{2}{*}{Problem} & \multirow{2}{*}{Metric}
& \multicolumn{2}{c|}{TR--Newton}
& \multicolumn{2}{c|}{TR--SR1}
& \multicolumn{2}{c|}{TR--BFGS}
& \multicolumn{2}{c@{}}{TR--DFP} \\
\cmidrule(lr){3-4} \cmidrule(lr){5-6} \cmidrule(lr){7-8} \cmidrule(l){9-10}
& & CG & Cauchy & CG & Cauchy & CG & Cauchy & CG & Cauchy \\
\midrule
\endfirsthead

\multicolumn{10}{c}{\tablename\ \thetable{} — continued}\\
\toprule
\multirow{2}{*}{Problem} & \multirow{2}{*}{Metric}
& \multicolumn{2}{c|}{TR--Newton}
& \multicolumn{2}{c|}{TR--SR1}
& \multicolumn{2}{c|}{TR--BFGS}
& \multicolumn{2}{c@{}}{TR--DFP} \\
\cmidrule(lr){3-4} \cmidrule(lr){5-6} \cmidrule(lr){7-8} \cmidrule(l){9-10}
& & CG & Cauchy & CG & Cauchy & CG & Cauchy & CG & Cauchy \\
\midrule
\endhead

\midrule \multicolumn{10}{r}{Continued on next page} \\
\endfoot

\bottomrule
\\
\caption{\textbf{Trust region evaluation.} Performance of trust region algorithms across 11 problems of varying geometry and dimension. The best value for each metric is bolded. Runtimes were measured on CPU.}
\endlastfoot

\multirow{5}{*}{Quad\_A}
  & Iterations  & \textbf{6} & 53 & 24 & 52 & 28 & 35 & 41 & 32 \\
  & Func Evals  & \textbf{19} & 160 & 73 & 157 & 85 & 106 & 124 & 97 \\
  & Grad Evals  & \textbf{7} & 54 & 25 & 53 & 29 & 36 & 42 & 33 \\
  & Time (s)    & \textbf{0.01} & 0.03 & \textbf{0.01} & \textbf{0.01} & \textbf{0.01} & \textbf{0.01} & 0.02 & \textbf{0.01} \\
  & Converged?  & T & T & T & T & T & T & T & T \\

\midrule
\multirow{5}{*}{Quad\_B}
  & Iterations  & \textbf{10} & 1000 & 19 & 1000 & 59 & 1000 & 1000 & 1000 \\
  & Func Evals  & \textbf{31} & 3001 & 58 & 3001 & 178 & 3001 & 3001 & 3001 \\
  & Grad Evals  & \textbf{11} & 1001 & 20 & 1001 & 60 & 1001 & 1001 & 1001 \\
  & Time (s)    & \textbf{0.01} & 0.54 & \textbf{0.01} & 0.22 & 0.02 & 0.22 & 0.44 & 0.25 \\
  & Converged?  & T & F & T & F & T & F & F & F \\

\midrule
\newpage
\multirow{5}{*}{Quad\_C}
  & Iterations  & \textbf{9} & 59 & 29 & 59 & 35 & 43 & 47 & 52 \\
  & Func Evals  & \textbf{28} & 178 & 88 & 178 & 106 & 130 & 142 & 157 \\
  & Grad Evals  & \textbf{10} & 60 & 30 & 60 & 36 & 44 & 48 & 53 \\
  & Time (s)    & 0.44 & 2.94 & 0.08 & \textbf{0.06} & 0.21 & 0.23 & 0.31 & 0.28 \\
  & Converged?  & T & T & T & T & T & T & T & T \\

\midrule
\multirow{5}{*}{Quad\_D}
  & Iterations  & \textbf{49} & 1000 & 195 & 1000 & 269 & 1000 & 1000 & 1000 \\
  & Func Evals  & \textbf{148} & 3001 & 586 & 3001 & 808 & 3001 & 3001 & 3001 \\
  & Grad Evals  & \textbf{50} & 1001 & 196 & 1001 & 270 & 1001 & 1001 & 1001 \\
  & Time (s)    & 2.50 & 51.03 & \textbf{0.50} & 1.02 & 1.98 & 5.37 & 7.60 & 5.69 \\
  & Converged?  & T & F & T & F & T & F & F & F \\

\midrule
\multirow{5}{*}{Quartic\_A}
  & Iterations  & 3 & 3 & 3 & 3 & 3 & 3 & 3 & 3 \\
  & Func Evals  & 10 & 10 & 10 & 10 & 10 & 10 & 10 & 10 \\
  & Grad Evals  & 4 & 4 & 4 & 4 & 4 & 4 & 4 & 4 \\
  & Time (s)    & 0.00 & 0.00 & 0.00 & 0.00 & 0.00 & 0.00 & 0.00 & 0.00 \\
  & Converged?  & T & T & T & T & T & T & T & T \\

\midrule
\multirow{5}{*}{Quartic\_B}
  & Iterations  & 12 & 12 & 25 & 14 & \textbf{11} & 14 & 86 & 14 \\
  & Func Evals  & 37 & 37 & 76 & 43 & \textbf{34} & 43 & 259 & 43 \\
  & Grad Evals  & 13 & 13 & 26 & 15 & \textbf{12} & 15 & 87 & 15 \\
  & Time (s)    & 0.01 & 0.01 & 0.01 & \textbf{0.00} & \textbf{0.00} & \textbf{0.00} & 0.03 & \textbf{0.00} \\
  & Converged?  & T & T & T & T & T & T & T & T \\

\midrule
\multirow{5}{*}{Rosen\_A}
  & Iterations  & \textbf{30} & 1000 & 147 & 1000 & 53 & 1000 & 51 & 1000 \\
  & Func Evals  & \textbf{91} & 3001 & 442 & 3001 & 160 & 3001 & 154 & 3001 \\
  & Grad Evals  & \textbf{31} & 1001 & 148 & 1001 & 54 & 1001 & 52 & 1001 \\
  & Time (s)    & \textbf{0.01} & 0.34 & 0.03 & 0.18 & \textbf{0.01} & 0.19 & \textbf{0.01} & 0.19 \\
  & Converged?  & T & F & T & F & T & F & T & F \\

\midrule
\multirow{5}{*}{Rosen\_B}
  & Iterations  & \textbf{4} & 40 & 81 & 39 & 1000 & 55 & 312 & 46 \\
  & Func Evals  & \textbf{13} & 121 & 244 & 118 & 3001 & 166 & 937 & 139 \\
  & Grad Evals  & \textbf{5} & 41 & 82 & 40 & 1001 & 56 & 313 & 47 \\
  & Time (s)    & 0.02 & 0.23 & 0.02 & \textbf{0.01} & 0.30 & \textbf{0.01} & 0.11 & \textbf{0.01} \\
  & Converged?  & T & T & T & T & F & T & T & T \\

\midrule
\multirow{5}{*}{Exp\_A}
  & Iterations  & \textbf{12} & 535 & 18 & 287 & 17 & 75 & 43 & 359 \\
  & Func Evals  & \textbf{37} & 1606 & 55 & 862 & 52 & 226 & 130 & 1078 \\
  & Grad Evals  & \textbf{13} & 536 & 19 & 288 & 18 & 76 & 44 & 360 \\
  & Time (s)    & 0.01 & 0.46 & \textbf{0.00} & 0.06 & 0.01 & 0.02 & 0.01 & 0.08 \\
  & Converged?  & T & T & T & T & T & T & T & T \\

\midrule
\multirow{5}{*}{Exp\_B}
  & Iterations  & \textbf{12} & 521 & 18 & 181 & 17 & 330 & 43 & 119 \\
  & Func Evals  & \textbf{37} & 1564 & 55 & 544 & 52 & 991 & 130 & 358 \\
  & Grad Evals  & \textbf{13} & 522 & 19 & 182 & 18 & 331 & 44 & 120 \\
  & Time (s)    & 0.01 & 0.42 & \textbf{0.00} & 0.04 & \textbf{0.00} & 0.07 & 0.01 & 0.03 \\
  & Converged?  & T & T & T & T & T & T & T & T \\

\midrule
\multirow{5}{*}{Genhumps}
  & Iterations  & 87 & 225 & 58 & 114 & 55 & \textbf{47} & 1000 & 1000 \\
  & Func Evals  & 262 & 676 & 175 & 343 & 166 & \textbf{142} & 3001 & 3001 \\
  & Grad Evals  & 88 & 226 & 59 & 115 & 56 & \textbf{48} & 1001 & 1001 \\
  & Time (s)    & 0.19 & 0.49 & 0.03 & 0.06 & 0.03 & \textbf{0.02} & 0.66 & 0.45 \\
  & Converged?  & T & T & T & T & T & T & F & F \\

\end{longtable}

\section{Future Work}

We plan to extend optHIM to handle constrained continuous optimization.  This will involve integrating techniques such as interior‐point, augmented Lagrangian, and active‐set methods, all built on our existing line search and trust region framework.

Scaling our methods to very large problems—including deep neural networks—poses new challenges.  In particular, we will explore custom autograd hooks and Hessian‐vector approximations that stream gradient and curvature information efficiently through complex computational graphs.

Finally, we will replicate our CPU‐based experiments on GPU hardware.  This study will assess whether the relative performance trends we observed hold when leveraging parallelism and specialized kernels, and will guide further optimizations for high-throughput environments.

\bibliographystyle{splncs04}
\bibliography{refs}

\begin{thebibliography}{1}
\providecommand{\url}[1]{\texttt{#1}}
\providecommand{\urlprefix}{URL }
\providecommand{\doi}[1]{https://doi.org/#1}

\bibitem{ConnGouldToint2000}
Conn, A.R., Gould, N.I.M., Toint, P.L.: Trust Region Methods. Society for Industrial and Applied Mathematics, Philadelphia, PA (2000)

\bibitem{Davidon1959}
Davidon, W.C.: Variable metric method for minimization. IBM Journal of Research and Development  \textbf{3}(1),  76--85 (1959)

\bibitem{FletcherPowell1963}
Fletcher, R., Powell, M.J.D.: A rapidly convergent descent method for minimization. The Computer Journal  \textbf{6}(2),  163--168 (1963)

\bibitem{Nocedal1980}
Nocedal, J.: Updating quasi‐newton matrices with limited storage. Mathematics of Computation  \textbf{35}(151),  773--782 (1980)

\end{thebibliography}

\end{document}